# Acceleration Profiles and Processing Methods for Parabolic Flight

Christopher E. Carr[1,2,*], Noelle C. Bryan[1], Kendall N. Saboda[1], Srinivasa A. Bhattaru[3], Gary Ruvkun[2], Maria T. Zuber[1]

[1] Massachusetts Institute of Technology, Department of Earth, Atmospheric and Planetary Sciences, Cambridge, MA, USA

[2] Massachusetts General Hospital, Department of Molecular Biology, Boston, MA, USA

[3] Massachusetts Institute of Technology, Department of Aeronautics and Astronautics, Cambridge, MA, USA

*Correspondence: 77 Massachusetts Ave Room 54-418, Cambridge MA 02138, USA. chrisc@mit.edu, +1-617-253-0786.

**Running Title:** Parabolic Flight Acceleration Data & Methods

Supported by NASA award NNX15AF85G and by the MIT Media Lab Space Exploration Initiative.

**Abstract**
Parabolic flights provide cost-effective, time-limited access to "weightless" or reduced gravity conditions experienced in space or on planetary surfaces, e.g. the Moon or Mars. These flights facilitate fundamental research – from materials science to space biology – and testing/validation activities that support and complement infrequent and costly access to space. While parabolic flights have been conducted for decades, reference acceleration profiles and processing methods are not widely available – yet are critical for assessing the results of these activities. Here we present a method for collecting, analyzing, and classifying the altered gravity environments experienced during a parabolic flight. We validated this method using a commercially available accelerometer during a Boeing 727-200F flight with 20 parabolas. All data and analysis code are freely available. Our solution can be easily integrated with a variety of experimental designs, does not depend upon accelerometer orientation, and allows for unsupervised and repeatable classification of all phases of flight, providing a consistent and open-source approach to quantifying gravito-intertial accelerations (GIA), or *g* levels. As academic, governmental, and commercial use of space increases, data availability and validated processing methods will enable better planning, execution, and analysis of parabolic flight experiments, and thus, facilitate future space activities.

## Introduction

Specialized aircraft perform parabolic flight trajectories to produce varying gravito-inertial acceleration (GIA) environments, measured in *g*[1]. As the aircraft rises at a 45° pitch, the vertical velocity increases and occupants on board the aircraft will experience ~1.8 *g*. Halfway through this climb, the lift and thrust are reduced, and the aircraft and its occupants fall simultaneously, producing roughly 30 s of a 0 *g*, freefall environment[1]. At the end of the parabola, the aircraft regains altitude, returning the occupants to an increased GIA environment before repeating as many as 60 subsequent parabolas[1]. In addition to the 0 *g*, freefall environment, the trajectory of the parabolas may be modified to achieve reduced *g* levels experienced at the surface of the Moon or Mars (0.17 and 0.38 *g*, respectively).

Parabolic flights are cost effective, ground-based analogues that achieve variable *g* level environments that recreate conditions experienced during spaceflight[1,2]. Such analogues offer insight into multiple aspects of spaceflight, for example, human physiology[2-4], materials performance[5], and behavior of liquid propellants[6]. Parabolic flights also provide researchers of diverse backgrounds with otherwise unattainable access to a spaceflight environment, fostering future research interest in multiple aspects of space science[7-9]. Parabolic flights serve as valuable proving grounds for experimental efforts to maximize the research potential of the International Space Station[10,11] and to accommodate increasing interest in commercial spaceflight[12,13].

Here we address: 1) the limited availability of open access acceleration datasets containing parabolic flight profiles, as well as 2) methods for their analysis. Specifically, we suggest a standardized approach to provide detailed time frames and precise *g* level characterization during all phases of flight. We demonstrate this approach using data collected with a small (65 g) battery powered commercially-available accelerometer and vibration measurement system (Slam Stick X™, Mide Technology Corp.) capable of up to kHz sampling rates. All data and code are provided online to enable planning of parabolic flight experiments and refinement of data processing methods. Together, these tools and products reflect a comprehensive solution to characterize the impact of altered *g* level and vibrations during parabolic flight experiments.

## Results

Flight operations were conducted on November 17, 2017 onboard the G-Force One®, a Boeing 727-200F aircraft operated by Zero Gravity Corporation. Four sets of parabolas were performed with 5, 6, 4, and 5 parabolas respectively. The first set targeted, in order, Mars *g*, Mars *g*, Lunar *g*, 0 *g*, and 0 *g*. All other parabolas targeted 0 *g*.

Data were collected for 1.77 hours during all phases of flight from a Slam Stick X™ mounted in the rear of the 18 m x 3 m research section (**Fig. 1a-b**). Sampling rates were 5 kHz (piezoelectric vibration sensor), 411 Hz (DC acceleration), and 1 Hz (pressure and temperature). Temperature varied less than 1˚C and pressure showed a typical regulated profile and was highly stable during parabolas (**Supplemental Fig. 3**). Specific force was concentrated in the z axis as measured by root mean square (rms) values (0.0466, 0.0775, 1.0662 for x, y, and z axes respectively), consistent with the accelerometer orientation (**Fig. 1b**).

In any given experiment, one accelerometer orientation may be more appropriate than another. Thus, we based our phase of flight identification method on a measure that is independent of the accelerometer orientation: the Euclidean norm of the x, y, and z axes, which we hereafter refer to as the *g* level or *g*. Because this variable is a positive scalar, it does not capture directional fluctuations in the gravity vector. Thus, when computing statistics, we recommend vector-based analysis. For example, estimating the mean *g* level during a 0 *g* parabola should be done by averaging the x, y, and z components and then computing the norm.

The expected value of the *g* level is unity on Earth under non-accelerated conditions; as a verification of our accelerometer calibration, we found the norm under lab bench conditions (14.2 s recording) to be 0.9840 (rms) and 0.9840 ± 0.0055 (mean ± standard deviation). This is a lower bound when vibration or specific force other than that caused by gravity is present, consistent with the rms value (1.07) observed during flight.

To facilitate parabola identification, acceleration data (**Fig. 1c**) were filtered using a zero-delay, low-pass filter (**Fig. 1d**; see Methods).

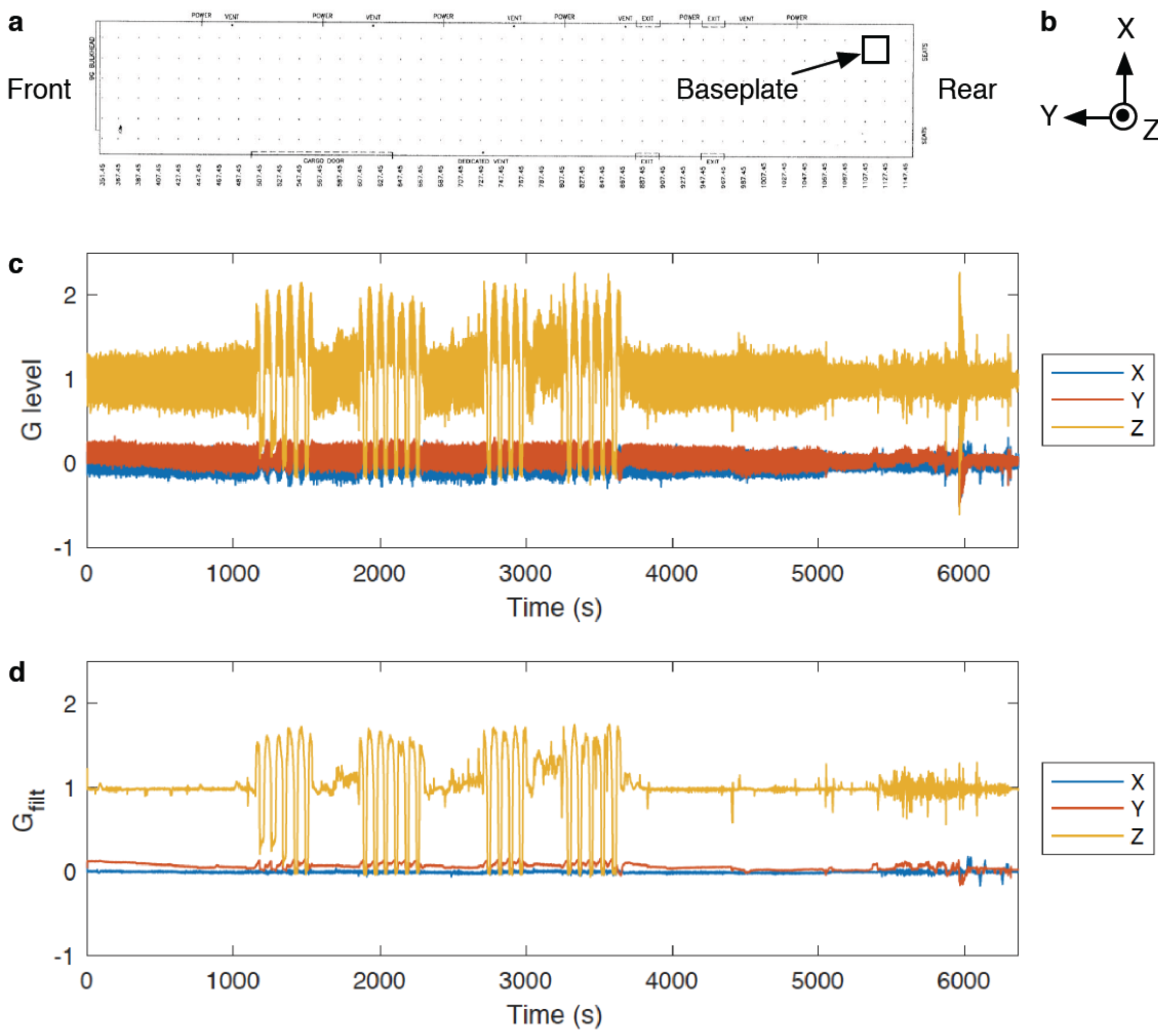

**Fig. 1. Acceleration profile of a parabolic flight.** (**a**) Research section of aircraft with baseplate location during flight; dots are separated by 50.8 cm. (**b**) Orientation of accelerometer on baseplate in panel A. (**c**) Raw DC accelerometer data. (**d**) Measured accelerations after low-pass filtering.

*Filter Optimization*

To optimize the filter, we selected a Half Power Frequency (HPF) based on the *g* level power spectral density (PSD, **Fig. 2a**). To select the HPF, we examined the cumulative sum of the PSD (**Fig. 2b**), which revealed a sharp increase in power above 0.01 Hz. We chose this value (HPF = 0.01 Hz) to maximize the low frequency content of the filtered data while rejecting as much spectral power from higher frequencies as possible. As an example, filtering at HPF = 0.01 Hz preserves parabola dynamics, while filtering at HPF = 0.001 Hz does not (**Fig. 2c**). Our selected value provides the smoothest data for identifying parabolas while still accurately representing *g* level transitions. A manual procedure identified similar values, e.g., adjusting the Half Power Frequency (HPF) toward DC until the rapid transitions between *g* levels showed systematic bias, then setting the HPF to 10X this value, also gave HPF = 0.01 Hz.

**a**

**b**

**c**

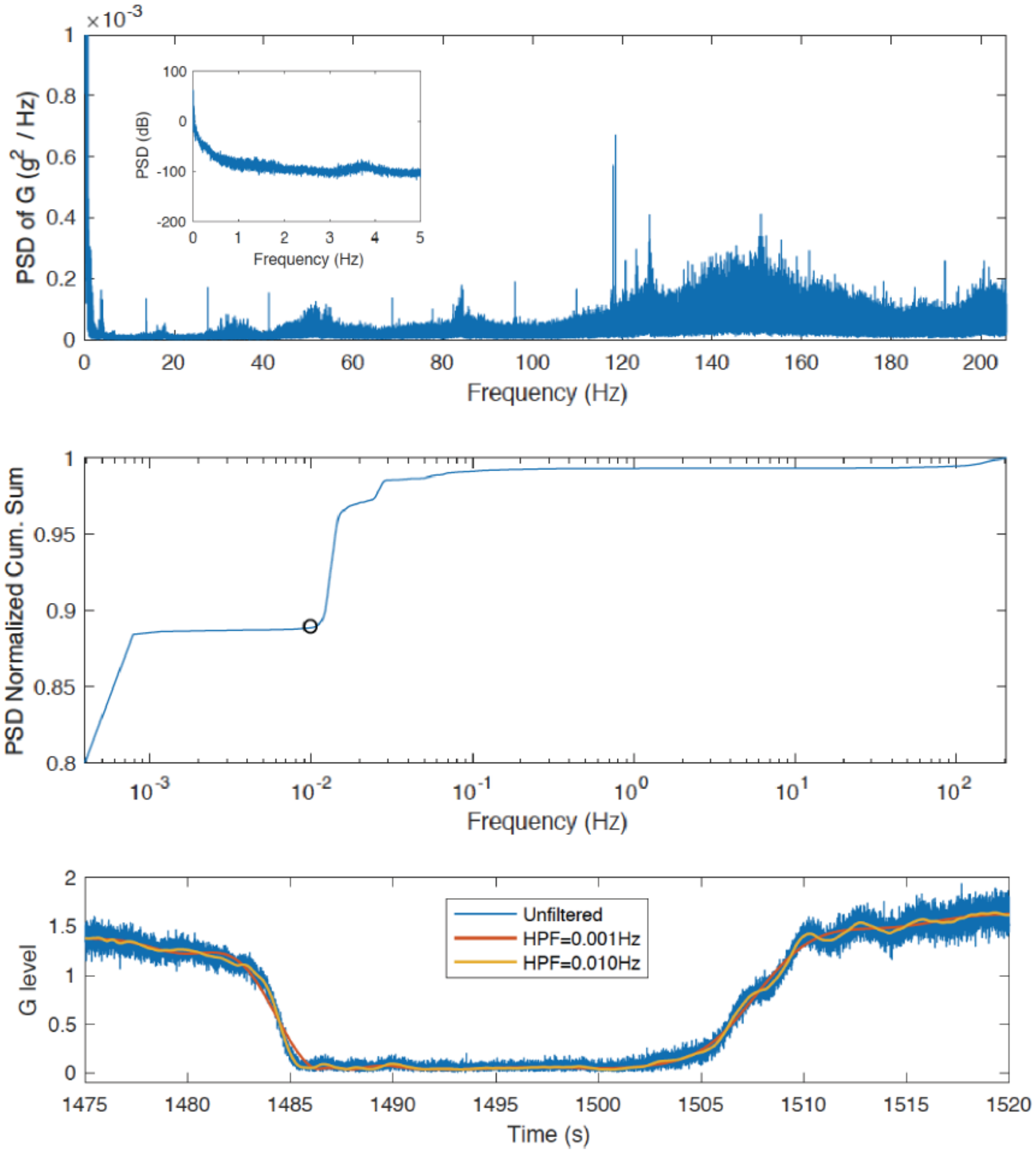

**Fig. 2. Power Spectral Density (PSD) of *g* level enables rational selection of filter Half Power Frequency.** (**a**) PSD estimated using Welch's method. Near DC frequencies hold much of the spectral power (inset). (**b**) Cumulative sum of PSD normalized to unity with selected filter HPF of 0.01 Hz (circle). (**c**) The *g* level filtered with HPF (0.01 Hz, orange) better matches the unfiltered *g* level (blue) than does a 10x lower frequency HPF (0.001 Hz, red).

Filtering reduced the root mean square specific force in the lateral (x) direction but little in other directions (0.0161, 0.0681, 1.0623 for x, y, z respectively), consistent with low frequency aircraft accelerations mainly in pitch. The *g* level was near unity during periods of relative calm (**Supplemental Fig. 1a-b**), including the first 1000 seconds of data collected during largely straight and level flight (rms 0.9919 and 0.9856, raw and filtered, respectively). This unfiltered estimate is 0.8% higher than under lab bench conditions, and both are consistent with accurate sensor calibration at DC to lower than 2% error, based on the factory calibration.

*Parabola Identification & Phase of Flight Classification*

We identified parabolas using change point detection[14,15]. Conceptually, this process finds the point for which a statistical property (e.g., mean), has minimum total residual error summed across two groups, e.g. before and after the change point. Here residual error is the difference between an observed value and the statistical property for the group.

Change point detection was first applied to the filtered *g* level $g_{filt}$ to identify differences in mean *g* levels. A known number of change points was specified based on the parabola number within each set, e.g. 2 times the number of parabolas, plus two additional transitions (first pre-parabola pull up; last post-parabola pull up) for each set of parabolas. In our case sets of 5, 6, 4, and 5 (20 total) parabolas become 12, 14, 10, and 12 change points.

This total number of change points (48) is supplied to the algorithm (see methods), which identified all *g* level change points in an unsupervised manner (**Fig. 3a-b**). We desire to break down the flight into regions of stable *g* levels. Thus, for each change point, we used a secondary change point detection to identify differences in the slope of the *g* level vs. time curves.

Data within 10 seconds of each change point was subjected to this secondary change point detection using a linear slope metric. This step successfully segmented the flight into regions of rapid "transition" (indicated by dotted lines, **Fig. 3c-e**) and more stable regimes. This resulted in 97 flight periods (2X the number of change points + 1).

Classification of non-"transition" flight periods into "parabola," "hypergravity," and "other" (which includes straight and level flight as well as standard rate turns) was then performed, first by categorizing any periods with duration > 100 s as "other", then by segmenting data according to *g* level ("parabola" ≤ 0.9 *g*, 0.9 < "other" ≤ 1.1 *g*, "hypergravity > 1.1 *g*). Despite its simplicity, this classifier achieved good separation between classes.

Parabola durations (mean ± s.d.) were 19.5 ± 1.4 s (0 *g*, *N*=17, range 17 to 24 s), 23.7 s (Lunar *g*; *N*=1), and 28.9 ± 0.7 s (Mars *g*; *N*=2). The *g* levels achieved were 0.041 ± 0.005 *g* (0 *g*) and 0.159 *g* (lunar *g*). Both Mars parabolas achieved 0.356 *g*, indicating high consistency between parabolas targeting similar *g* levels. Higher *g* levels were significantly associated with longer-duration parabolas (**Fig. 4a; Supplemental Fig. 2a**), although not when lunar and Mars data were excluded (**Supplemental Fig. 2b)**.

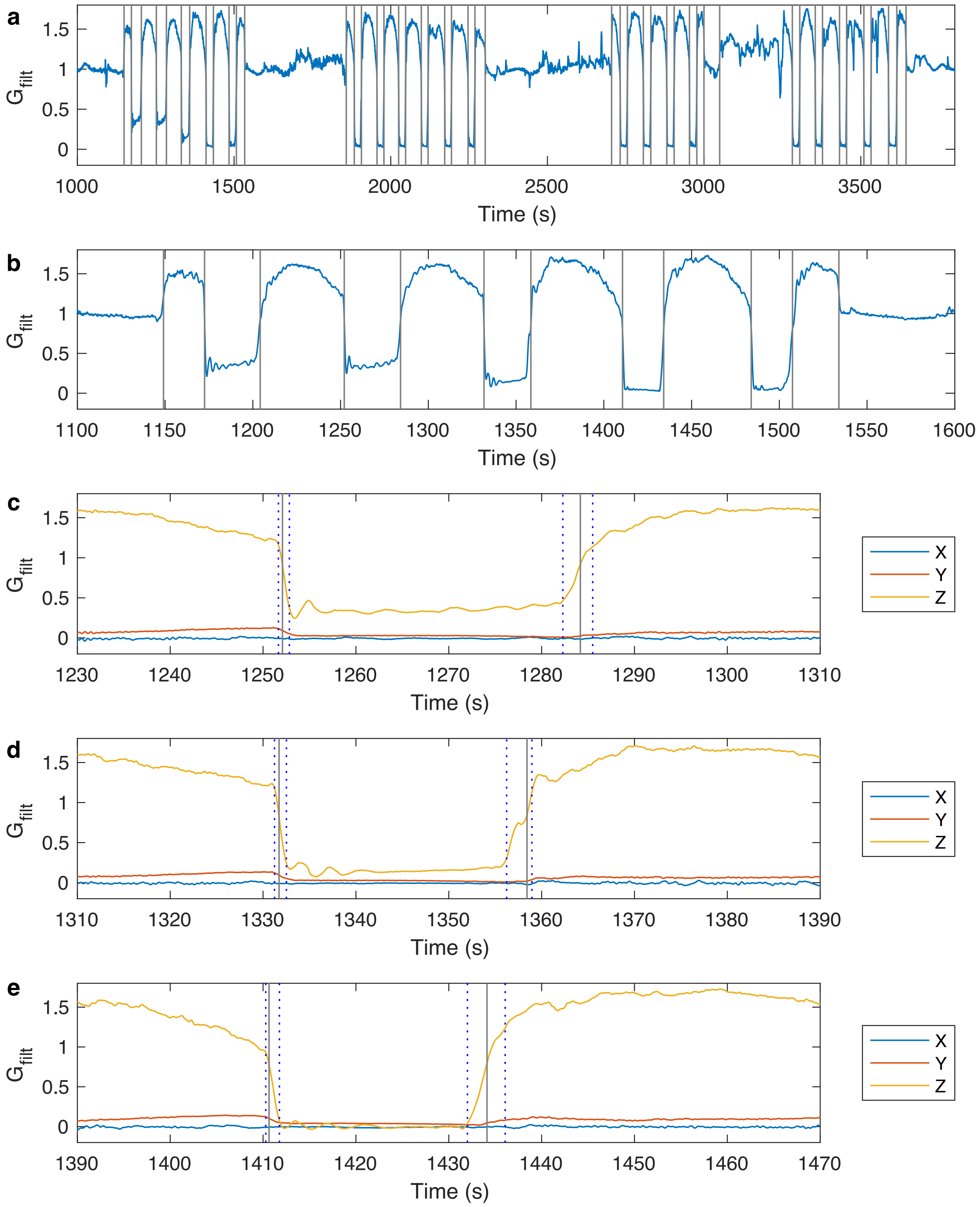

**Fig. 3. Parabola identification using change point detection.** (**a**) Change points (vertical lines) for mean *g* levels as measured by $g_{filt}$. (**b**) Zoom of (**a**) for the first set of parabolas. (**c-d**) Second level linear change points (vertical dotted lines) define transition regions for a Mars (**c**), a lunar (**d**), and a 0 *g* (**e**) parabola.

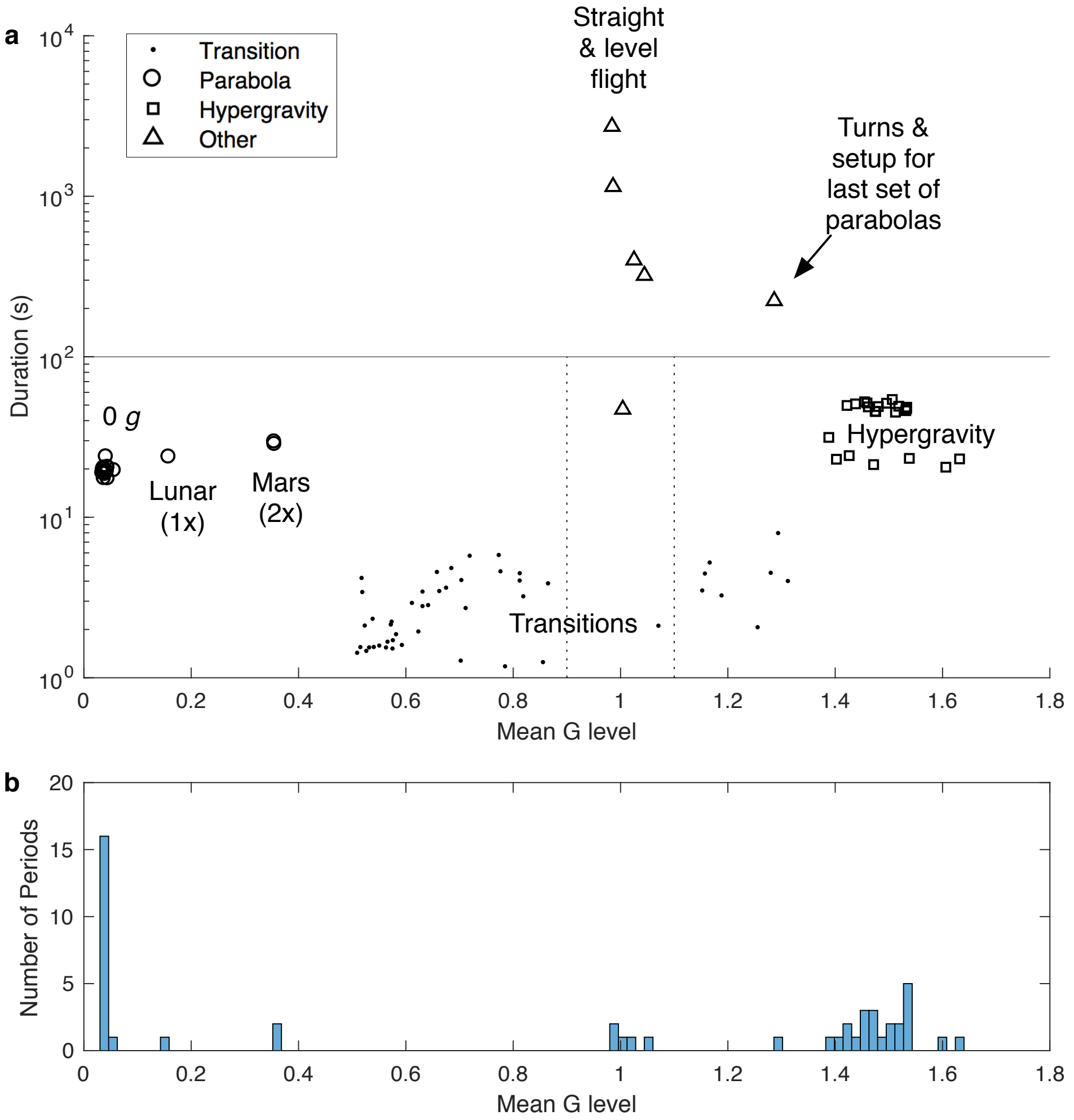

**Fig. 4. Phases of flight classification of non-transition periods.** (**a**) Non-transition events were classified as duration < 100 seconds (horizontal line) and by mean *g* level (dotted lines) into "parabola," "hypergravity," and "other." (**b**) Histogram by mean *g* level, estimated as norm of mean *g* vector during each period. All phases of flight were unambiguously classified. The subset of "hypergravity" periods with lower duration (bottom right) were identified as those at the start or end of a set of parabolas. The MATLAB script executing all data processing is freely available and writes the final set of flight periods to a delimited file for easy downstream processing.

## Discussion

Parabolic flights provide the opportunity to perform simulated space research in a cost-effective manner. However, there are currently no widely available datasets describing the altered *g*, nor the time scales encountered in such conditions, during parabolic flight. Lambot and Ord (2016) evaluated data from over 400 parabolic flights and assessed the quality of reduced *g* data sets. While considerable effort was dedicated to identifying the highest quality, low *g* time periods (with variations less than ±0.01 *g*) from these flights, neither the acquisition hardware, the raw data, nor the code implemented for analysis, are currently available to the public[16].

Here, we provide: 1) a commercially-available hardware solution for data acquisition, 2) raw and calibrated data for all phases of flight, 3) data analysis methodology that is independent of accelerometer orientation, and 4) characterization of *g* levels and durations achieved for 20 parabolas. Additionally, the code implementing our methodology to categorize all phases of flight and characterize *g* levels and durations of parabolas is publicly available in order to facilitate future parabolic flight research.

Our selected data acquisition system, the Slam Stick X™, offers a compact, flexible, low-power, high resolution solution for acceleration and vibration monitoring. Although here we have not assessed aircraft vibration beyond a cursory analysis of the DC accelerometer-measured power spectral density at low frequencies, 5 kHz data from the piezoelectric accelerometer is available in our online dataset. Alternative data acquisition systems include many commercial off the shelf (COTS) accelerometers, as well as the NASA Suborbital Flight Environment Monitor (SFEM)[16]. There may be potential benefits of using the SFEM, although the Slam Stick X™ offers comparable or longer recording time, DC and piezoelectric accelerometers (enabling both *g* level and high frequency vibration measurements), higher sampling frequencies, a wider operating temperature range (-40 ˚C to +80˚C) and much lower (>10X) mass and volume. Another COTS option is the Lansmont 3X90, although the Slam Stick X™ specifications provide benefits in several areas (size, mass, sampling rates, and temperature range).

One consideration for parabolic flight experiments is that the GIA environment is not constant across the aircraft. In some cases, it may be adequate to have a single reference flight profile to be used by multiple experiment teams. However, some applications may be better served through measurement of the local GIA environment of a given experimental apparatus. Here, the small size of our solution facilitates direct incorporation into a payload, as well as placement in the desired location or orientation.

When selecting a data acquisition solution, it is also important to consider how the mounting of the accelerometer itself may impact the frequency response; in our case, use of double-sided sticky tape (see Methods) represents both an extremely practical and low bias option, enabled by the low device mass. Because no additional materials separate the accelerometer from the aircraft, there is no need to correct for the frequency response of the mounting interface.

Some limitations are inherent in our study, which focused solely on one flight and 20 parabolas. If analyzing multiple flights, with parabolas performed under more varied conditions, it is possible a slightly more complex classification strategy might be required; however, based on the wide separation between “parabola,” “hypergravity,” and “other” classes, this is not expected to present a significant challenge to standard unsupervised classification approaches (e.g., *k*-means).

The relatively short time periods for 0 *g*, freefall environments require strategic planning to precisely determine the sampling rate and types of data to be collected. For instance, if instrument performance or certain tasks are to be evaluated during reduced *g* conditions, then results must be attained quickly to avoid being influenced by the successive transitions and high *g* conditions. However, the cumulative experimental time of 6.8 minutes over 20 parabolas in reduced *g* environments produced from a typical parabolic flight should provide adequate time for properly designed experiments and a means to prove functionality for technology intended for deployment in space, such as onboard the International Space Station (ISS).

Due to the limited availability and high cost of actual space environments, it is imperative that we continue to utilize parabolic flights as a means to simulate space – and to understand the accuracy and limitations of this modality. By making our data and methods available we hope to enable others to better plan, execute, and analyze parabolic flight experiments, and thus, to help facilitate future space activities.

## Methods

### *Device Selection*

The Slam Stick X (Mide Technology Corp., www.mide.com) was selected based on its size (76 mm x 30 mm x 15 mm), low mass (65 g), integrated battery, manual and USB interfaces, and combination DC (Analog Devices ADXL345) and piezoelectric (TE 832M1) accelerometers to enable accuracy at both low (e.g. down to 0 Hz) and high frequencies (up to 20 kHz sample frequency). The aluminum body was selected to provide improved high frequency response. Additional integrated sensors included temperature and pressure (NXP MPL3115) and control pad temperature and pressure (TE MS8607).

### *Device Mounting and Data Acquisition*

Zero Gravity Corporation (ZGC) utilizes a standard system of mounting hardware to the aircraft structure consisting of a baseplate (61 cm x 61 cm x 1.27 cm aluminum plate, e.g. McMaster Carr 86825K25) bolted to the aircraft structure using four clearance holes at the corners of a square with 50.8 cm (20 in) sides, centered on the baseplate. Washers (McMaster Carr 92503A230) were used for mounting in combination with AN-6 steel bolts (3/8 inch) provided by ZGC. The Slam Stick X was mounted to a standard baseplate with double-sided sticky tape (3M 950), which is the preferred mounting method due to its vibration frequency response (near unity) and robustness: this method has previously been validated during vibration testing at over 75 g at 1 kHz[1]. The Slam Stick X was configured using Slam Stick Lab 1.8. Acquisition was initiated and terminated manually using the control pad on the device.

### *Data Processing*

The raw IDE file generated by the Slam Stick X was converted to a MAT (MATLAB, The Mathworks, Natick, MA) file using the raw2mat.exe command line utility (Mide Technology Corp.). Upon loading data into MATLAB R2017a, the raw sensor readings were calibrated using the factory calibration. Note that data calibration and export functions can also be performed directly using Slam Stick Lab. Here we focus on data from the DC accelerometer. First, the power spectral density was computed using the MATLAB *pwelch()* function using default parameters. For parabola identification, we first filtered the raw data using a zero-phase 12$^{th}$ order Butterworth filter using the *designfilt()* function (see text for details). Next, we utilized change point detection[14,15] as implemented by the MATLAB *FindChangePts()* function as described in the text. Regression of parabola *g* level on duration was performed using the MATLAB *fitlm()* function. Confidence intervals were determined using the MATLAB *coefCI()* function.

[1] Mide Technology Corp., Slam Stick User Manual, Version 2.0

*Code Availability*
The MATLAB scripts implementing our analysis are available at:
https://github.com/CarrCE/zerog

*Data Availability*
Raw and calibrated data are available via the Open Science Framework at:
https://osf.io/nk2w4/

**Acknowledgements:** This work was supported by NASA award NNX15AF85G and by the MIT Media Lab Space Exploration Initiative.

**Contributions:** C.E.C. designed the experiment, C.E.C., K.S., S.A.B. and N.B. built and tested the hardware, N.B. and M.Z. collected the data, C.E.C. and N.B. processed the data. G.R. advised on the experiment design. C.E.C. and N.B. wrote, and all authors edited and approved, the paper. C.E.C. is the guarantor.

**Competing Interests:** The authors declare no conflict of interest.

**Supplementary Information:** Supplementary information (Supplementary Figures 1-3) are available below.

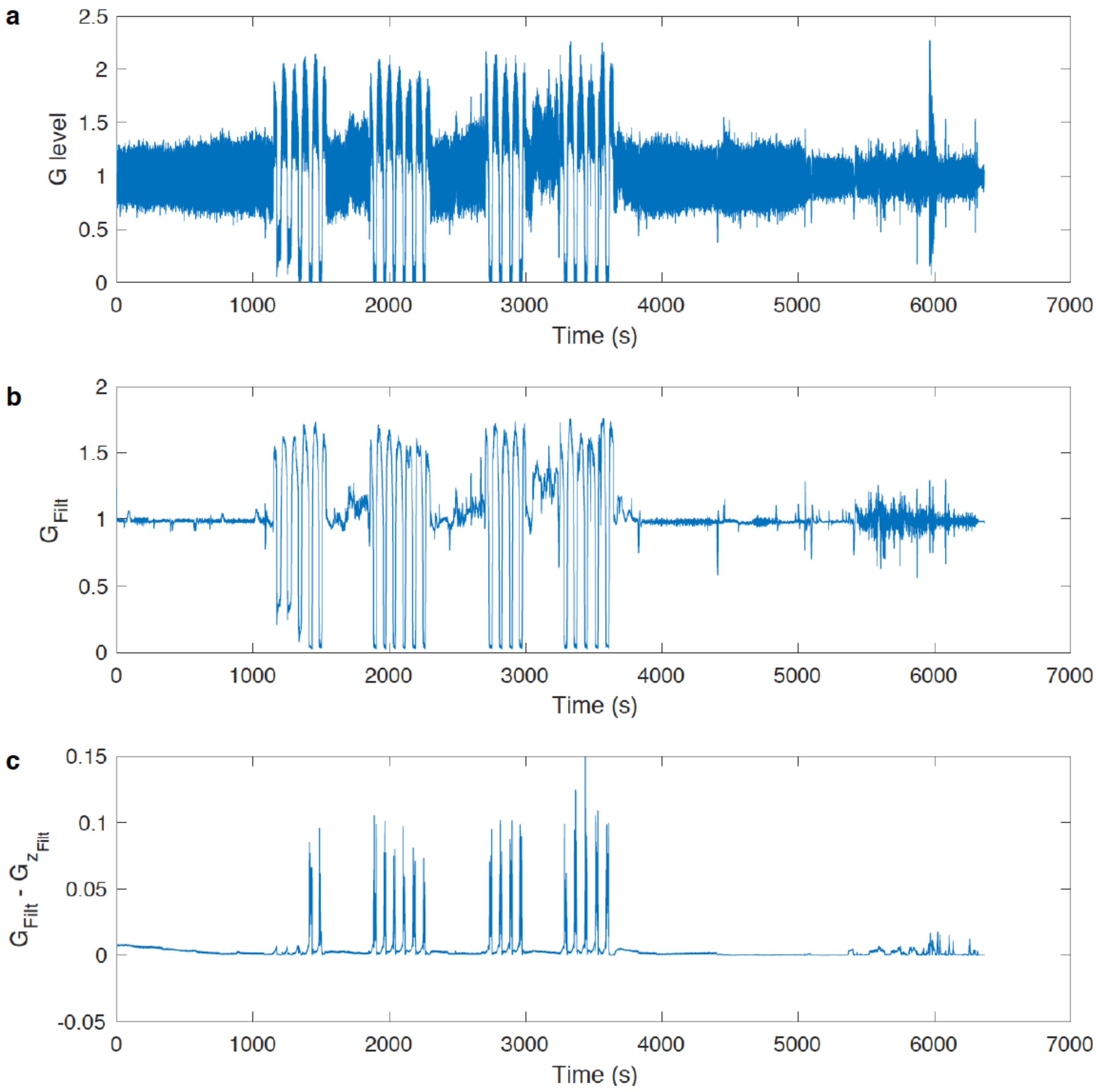

**Supplementary Fig. 1. Acceleration profile as measured by *g* level, the norm of the gravity vector.** (**a**) Unfiltered *g* level. (**b**) Low-pass filtered *g* level. (**c**) Difference between *g* level and magnitude of z-axis acceleration $g_z$. This difference is appreciable during 0 *g* parabolas but not Mars or lunar *g* parabolas, suggesting significant contributions to *g* from x and y axes during zero *g* parabolas. Accelerometer axes are as described in **Fig. 1b**.

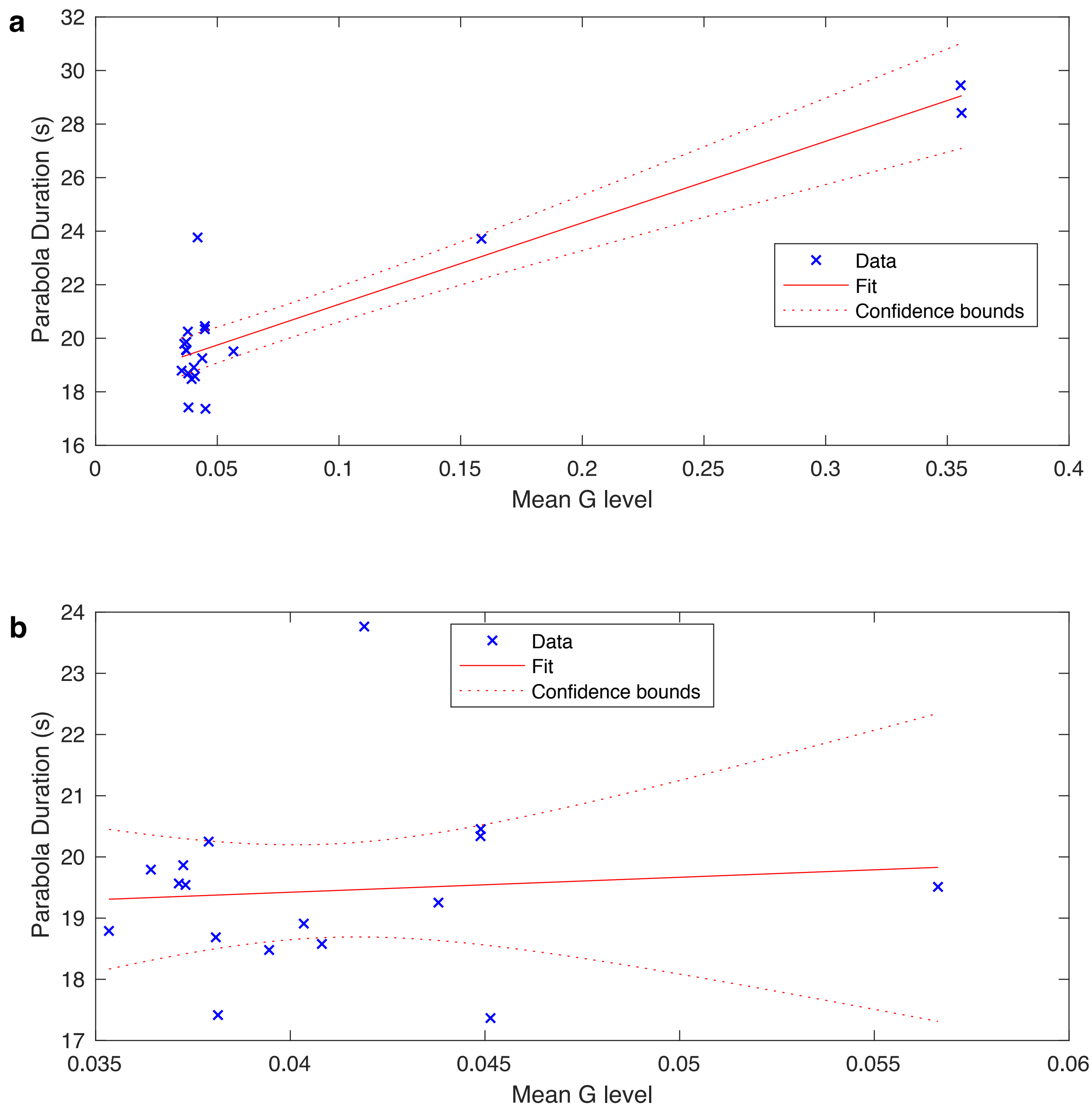

**Supplemental Fig. 2. Parabola duration as a function of *g* level.** (**a**) Regression of *g* level on parabola duration (*N*=20) yielded a highly significant relationship ($p<10^{-7}$). (**b**) However, a regression without the limited Lunar *g* (*N*=1) and Mars *g* (*N*=2) data was insignificant ($p$=0.737). Thus, care should be taken not to over-interpret the measured relationship.

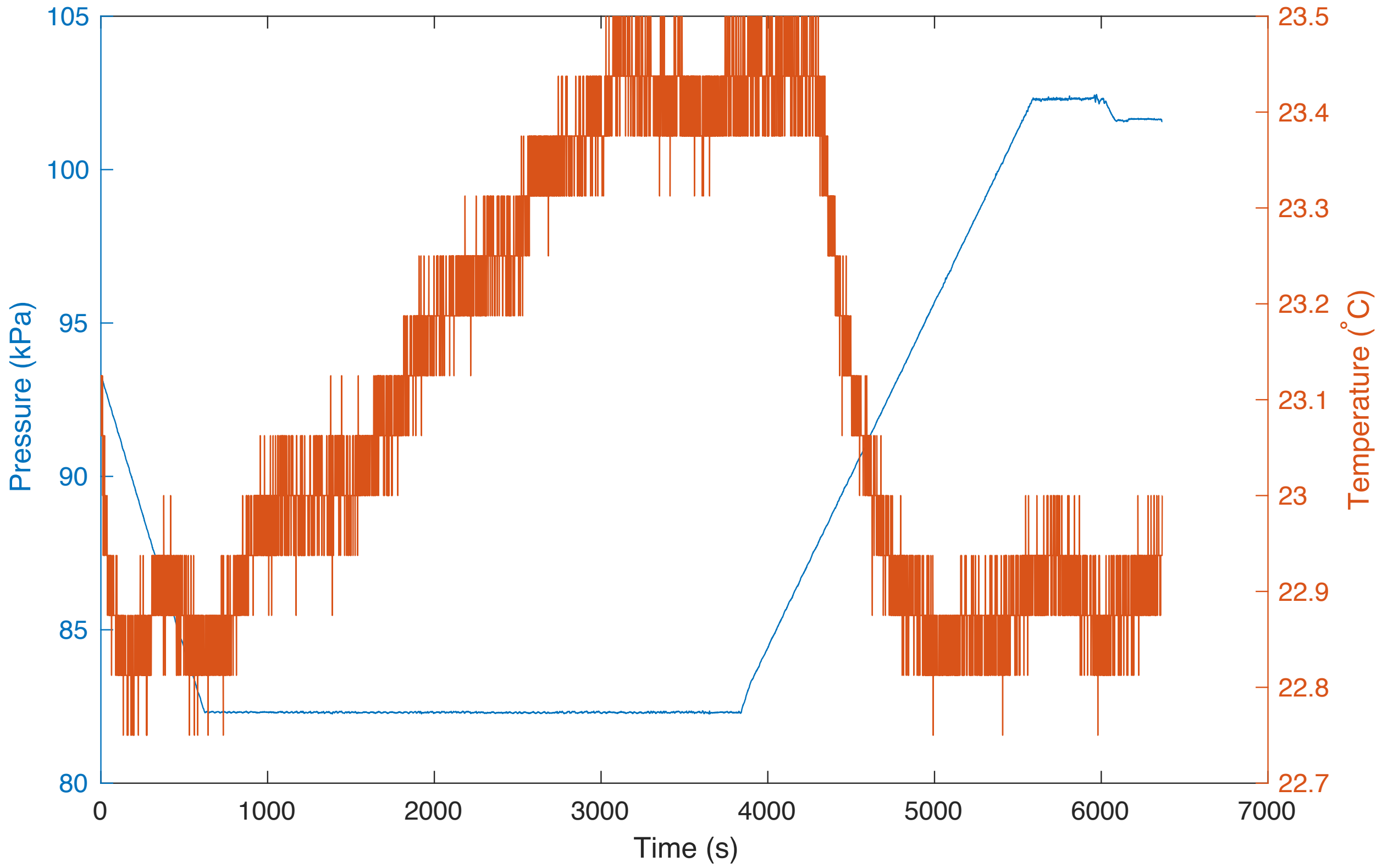

**Supplemental Fig. 3. Pressure and temperature during flight.** Pressure profile (blue line) reflects pressure altitude of ~1720 m established during flight until completion of parabolas. Measured temperature varied within a tight range around 23˚C.